\documentstyle[aps,prl,multicol,epsf,amssymb]{revtex}

\begin{document}
\bibliographystyle{prsty}
\draft

\title{Thermal Transport Measurements of Individual Multiwalled Nanotubes}

\author{P. Kim$^1$, L. Shi$^2$, A. Majumdar$^2$, P. L.
McEuen$^{1,3}$\cite{paul}}

\address{$^1$Department of Physics, University of California, Berkeley,
CA 94720}
\address{$^2$Department of Mechanical Engineering, University of California, Berkeley,
CA 94720}
\address{$^3$Division of Materials Sciences, Lawrence Berkeley National Laboratory, Berkeley,
CA 94720}

\maketitle

\begin{abstract}

The thermal conductivity and thermoelectric power of a single
carbon nanotube were measured using a microfabricated suspended
device. The observed thermal conductivity is more than 3000 W/K m
at room temperature, which is two orders of magnitude higher than
the estimation from previous experiments that used macroscopic mat
samples. The temperature dependence of the thermal conductivity of
nanotubes exhibits a peak at 320 K due to the onset of Umklapp
phonon scattering. The measured thermoelectric power shows linear
temperature dependence with a value of 80 $\mu$V/K at room
temperature.

\end{abstract}

\begin{multicols}{2}
\narrowtext


The thermal properties of carbon nanotubes are of fundamental
interest and also play a critical role in controlling the
performance and stability of nanotube devices \cite{[1]}. Unlike
electrical and mechanical properties, which have been studied at a
single nanotube level \cite{[2]}, the thermal properties of carbon
nanotubes have not been measured at a mesoscopic scale. The
specific heat, thermal conductivity and thermoelectric power (TEP)
of millimeter-sized mats of carbon nanotubes have been measured by
several groups
\cite{[3],[4],[5],[6],[7],[8],[9],[10],[11],[12],[13]}. Although
these studies have yielded a qualitative understanding of the
thermal properties of these materials, there are significant
disadvantages to these 'bulk' measurements for understanding
intrinsic thermal properties of a single nanotube.  One problem is
that these measurements yield an ensemble average over the
different tubes in a sample.  More importantly, in thermal
transport measurements such as thermal conductivity and TEP, it is
difficult to extract absolute values for these quantities due to
the presence of numerous tube-tube junctions.  These junctions are
in fact the dominant barriers to thermal transport in a mat of
nanotubes.

In this letter, we present the results of mesoscopic thermal
transport measurements of individual carbon nanotubes. We have
developed a microfabricated suspended device hybridized with
multiwalled nanotubes (MWNTs) to probe thermal transport free from
a substrate contact. The observed thermal conductivity of MWNT is
two orders of magnitude higher than the value found in previous
'bulk' measurements and is comparable to the theoretical
expectations.

Suspended structures were fabricated on a silicon nitride/silicon
oxide/silicon multilayer by electron beam and photo lithography
followed by metalizations and etching processes, which are
described elsewhere in detail \cite{[14]}. Fig.\ref{fig1}(a) shows
a representative device including two 10 $\mu$m $\times$ 10 $\mu$m
adjacent silicon nitride membrane (0.5 $\mu$m thick) islands
suspended with 200~$\mu$m long silicon nitride beams. On each
island, a Pt thin film resistor, fabricated by electron beam
lithography, serves as a heater to increase the temperature of the
suspended island. These resistors are electrically connected to
contact pads by the metal lines on the suspending legs. Since the
resistance of the Pt resistor changes with temperature
(Fig.\ref{fig1}(b)), they also serve as a thermometer to measure
the temperature of each island \cite{[15]}.

\setcounter{figure}{0}
\begin{figure}
\epsfxsize=65mm \centerline{\epsffile{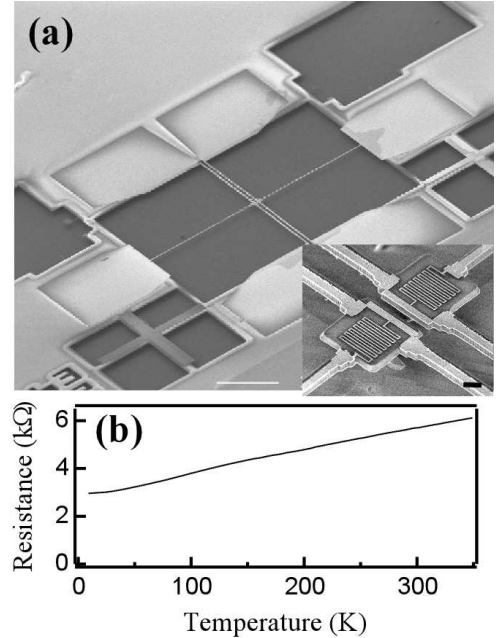}} \vskip 4mm
\caption{(a) A large scale scanning electron microscopy (SEM)
image of a microfabricated device. Two independent islands are
suspended by three sets of 250~$\mu$m long silicon nitride legs
with Pt lines that connect the micro-thermometer on the islands to
the bonding pads. Scale bar represents 100~$\mu$m.  (inset)
Enlarged image of the suspended islands with the Pt resistors. The
scale bar represents 1~$\mu$m. (b) The resistance of the Pt
resistor over the measured temperature ranges.} \label{fig1}
\end{figure} \vskip 0mm

\begin{figure}
\epsfxsize=95mm \centerline{\epsffile{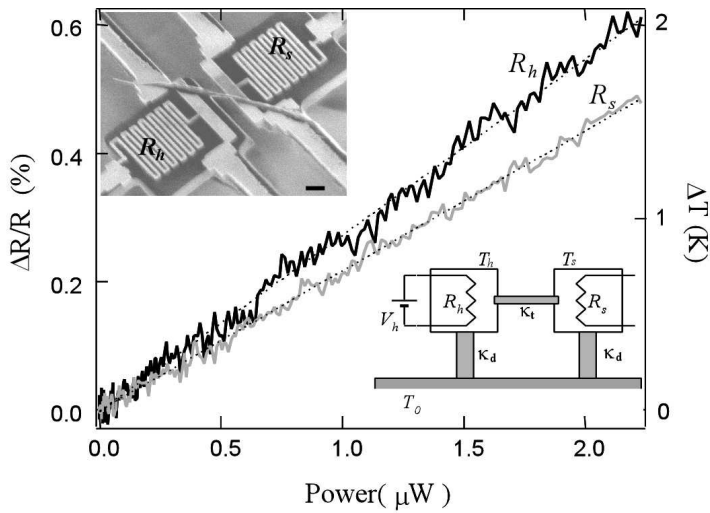}} \vskip 3mm
\caption{The change of resistance of the heater resistor ($R_h$)
and sensor resistor ($R_s$) as a function of the applied power to
the heater resistor. (upper inset) SEM image of the suspended
islands with a MWNT bundle across the device. The scale bar
represents 1~$\mu$m. (lower inset) A schematic heat flow model of
the device.} \label{fig2}\end{figure}

Once the suspended devices were fabricated, carbon nanotubes were
placed on the device and bridged the two suspended islands.
Mechanical manipulation similar to that used for the fabrication
of nanotube scanning probe microscopy tips \cite{[16]} was used to
place MWNTs on the desired part of the device.  This approach
routinely produces a nanotube device that can be used to measure
the thermal conductivity and TEP of the bridging nanotube segment.
Shown in Fig.\ref{fig2} (upper inset) is an example of such a
device. A small MWNT bundle forms a thermal path between two
suspended islands that are otherwise thermally isolated to each
other. A bias voltage applied to one of the resistors, $R_h$,
creates Joule heat and increases the temperature, $T_h$, of the
heater island from the thermal bath temperature $T_0$. Under
steady state, there is a heat transfer to the other island through
the nanotubes, and thus the temperature, $T_s$, of the resistor
$R_s$ also rises. Using a simple heat transfer model (lower inset
to Fig.\ref{fig2}), the thermal conductance of the connecting
nanotubes, $K_t$, and the suspending legs, $K_d$, can readily
estimated from: $T_h = T_0+\frac{K_d+K_t}{K_d(K_d+2K_t)}P$ and
$T_s =T_0+\frac{K_t}{K_d(K_d+2K_t)}P$, where $P$ is the Joule
power applied to the resistor $R_h$. Fig.\ref{fig2} shows the
temperature changes of each of the suspended islands connected by
the nanotubes as a function $P$. From the slopes of $R_s$ and
$R_h$ versus $P$, the thermal conductance of the bridging
nanotubes at the temperature $T_0$ can be computed using above
equations.

\vskip 5mm
 \vskip 0mm

\begin{figure}
\epsfxsize=85mm \centerline{\epsffile{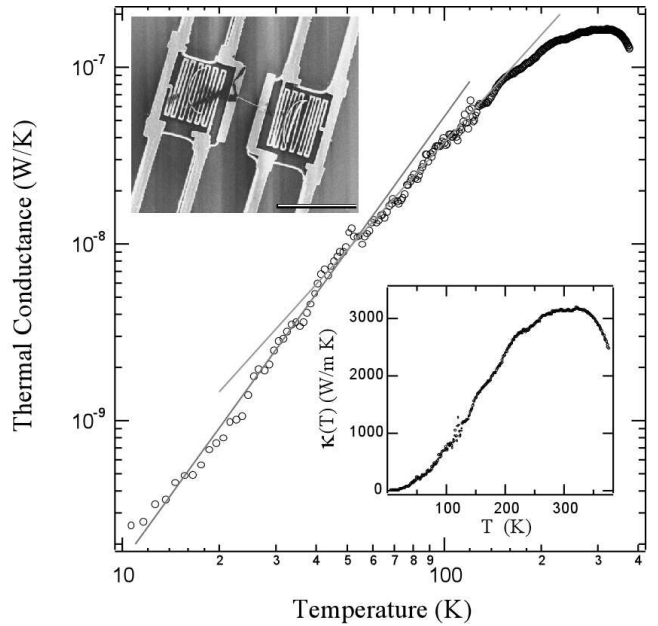}} \vskip 5mm
\caption{The thermal conductance of an individual MWNT of a
diameter 14 nm. The solid lines represent linear fits of the data
in a logarithmic scale at different temperature ranges. The slopes
of the line fits are 2.50 and 2.01, respectively. (lower inset)
Thermal conductivity of the MWNT. (upper inset) SEM image of the
suspended islands with the individual MWNT. The scale bar
represents 10~$\mu$m.} \label{fig3}
\end{figure} \vskip 0mm

We now turn to the experimental results of thermal conductance
measurement of a single MWNT. Fig.\ref{fig3} displays measured
thermal conductance of a single MWNT with a diameter 14 nm and the
length of the bridging segment 2.5~$\mu$m. The thermal conductance
was measured in a temperature range 8-370 K \cite{[17]}. It
increases by several orders of magnitude as the temperature is
raised, reaching a maximum of approximately
1.6$\times$10$^{-7}$~W/K near room temperature before decreasing
again at higher temperatures.

The measured thermal conductance includes the thermal conductance
of the junction between the MWNT and the suspended islands in
addition to the intrinsic thermal conductance of the MWNT itself.
From our separate study of scanning thermal microscopy on a self
heated MWNT \cite{[18]}, we have estimated the thermal conductance
of the junction at room temperature; the heat flow rate from a
unit length of the tube to metal electrode at a given unit
junction temperature difference was found to be $\sim$~0.5 W/m K.
Considering the contact length of the MWNT to the electrodes on
the islands is $\sim$1~$\mu$m, the junction thermal conductance is
$\sim$ 5$\times$10$^{-7}$~W/K at room temperature.  Since the
total measured thermal conductance is 1.6$\times$10$^{-7}$~W/K,
this suggests that the intrinsic thermal conductance of the tube
is the major part of measured thermal conductance.

To estimate thermal conductivity from this measured thermal
conductance, we have to consider the geometric factors of the MWNT
and the anisotropic nature of thermal conductivity.  The outer
walls of the MWNT that make good thermal contacts to a thermal
bath have more contribution in thermal transport than the inner
walls, and the ratio of axial to radial thermal conductivity may
influence the conversion of thermal conductance to thermal
conductivity. In the following, however, we simply estimate the
thermal conductivity neglecting the junction thermal conductance
and assuming a solid isotropic material to correct geometric
factors. This simplification implies that the thermal conductivity
reported in this letter is a lower bound of the intrinsic axial
thermal conductivity of a MWNT. Further study to analyze the
contribution of individual layers of MWNTs \cite{[19]} in the
thermal transport should elucidate this important issue in future.

Shown in inset to Fig.\ref{fig3} is the temperature dependent
thermal conductivity, $\kappa(T)$. This result shows remarkable
differences from the previous 'bulk' measurements as described
below. First, the room temperature value of $\kappa(T)$ is over
3000~W/m~K~\cite{[20]}, whereas the previous 'bulk' measurement on
a MWNT mat using the 3w method estimates only 20~W/m~K~\cite{[4]}.
Note that our observed value is also an order of magnitude higher
than that of aligned SWNT sample (250 W/m K) \cite{[7]} but
comparable to the recent theoretical expectation,
6000~W/m~K~\cite{[21]}. This large difference between single-tube
and 'bulk' measurements suggests that numerous highly resistive
thermal junctions between the tubes largely dominate the thermal
transport in mat samples. Second, $\kappa(T)$ shows interesting
temperature dependent behavior that was absent in 'bulk'
measurement. At low temperatures, 8~K$<T<$50~K, $\kappa(T)$
increases following a power law with an exponent 2.50. In the
intermediate temperature range (50~K$<T<150$~K), $\kappa(T)$
increases almost quadratically in $T$ (i.e., $\kappa(T)\sim T^2$).
Above this temperature range, $\kappa(T)$ deviates from quadratic
temperature dependence and has a peak at 320 K. Beyond this peak,
$\kappa(T)$ decreases rapidly.

We now seek to understand the physics behind the observed behavior
of the thermal conductivity. In a simple model \cite{[22]}, the
phonon thermal conductivity can be written as: $\kappa= \sum_p C_p
v_p l_p$, where $C_p$, $v_p$, and $l_p$ is the specific heat
capacity, phonon group velocity, and the mean free path of phonon
mode $p$. The phonon mean free path consists of two contributions:
$l^{-1}=l_{st}^{-1}+l_{um}^{-1}$ where $l_{st}$ and $l_{um}$ are
static and Umklapp scattering length, respectively. At low
temperatures, the Umklapp scattering freezes out, $l=l_{st}$, and
thus $\kappa(T)$ simply follows the temperature dependence of
$C_p$'s. For MWNTs, below the Debye temperature of interlayer
phonon mode, $\Theta_{\bot}$, $\kappa(T)$ has slight three
dimensional nature, and $\kappa(T) \sim T^{2.5}$ as observed in
graphite single crystals \cite{[23]}. As $T>\Theta_{\bot}$, the
interlayer phonon modes are fully occupied, and $\kappa(T) \sim
T^2$, indicative of the two dimensional nature of thermal
conduction in a MWNT \cite{[24]}. From this cross over behavior of
$\kappa(T)$, we estimate $\Theta_{\bot}= 50$~K. This value is
comparable to the value obtained by a measurement of specific heat
of MWNT~\cite{[4]}.

\begin{figure}[t] \epsfxsize=85mm
\centerline{\epsffile{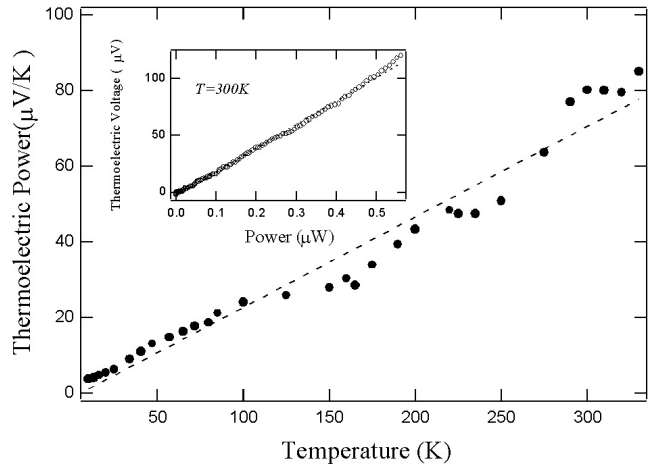}} \vskip 6mm \caption{The measured
thermoelectric power (solid circles) and a linear fit (broken
lines). (inset) The thermoelectric voltage versus the power
applied to the heater resistor at 300 K. The broken line
corresponds a linear fit.} \label{fig4}
\end{figure}

As $T$ increases further, the strong phonon-phonon Umklapp
scattering becomes more effective as higher energy phonons are
thermally populated. Once $l_{st} > l_{um}$, $\kappa(T)$ decreases
as $T$ increase due to rapidly decreasing $l_{um}$. At the peak
value of $\kappa(T)$, where $l_{st}\sim l_{um}$ ($T = 320$ K), we
can estimate the $T$-independent $l_{st}\sim 500$~nm for the MWNT.
Note that this value is an order of magnitude higher than previous
estimations from 'bulk' measurements~\cite{[7]} and is comparable
to the length of the measured MWNT (2.5 $\mu$m). Thus below room
temperature where the phonon-phonon Umklapp scattering is minimal,
phonons have only a few scattering events between the thermal
reservoirs, and the phonon transport is 'nearly' ballistic. This
remarkable behavior was not seen in the bulk experiments, possibly
due to additional extrinsic phonon scattering mechanisms such as
tube-tube interactions.

In addition to the probing the thermal conductivity, the same
suspended device can be used to measure the thermoelectric power
of the nanotube. The two independent electrodes that contact the
ends of MWNTs on each suspended island serve to measure the
electrical potential difference across the tube when joule heating
in $R_h$ creates a temperature gradient across the tube. Shown in
the inset to Fig.\ref{fig4} is the relation of the thermoelectric
voltage of the suspended MWNT to the joule heating in $R_h$. Since
the temperature difference can be computed from an independent
measurement of $R_h$ and $R_s$, the TEP of the single MWNT is
obtained from the slope of this curve. The observed temperature
dependent TEP of MWNTs (Fig.\ref{fig4}) shows a linear $T$
dependence, with room temperature value 80 $\mu$V/K. The linear
$T$ dependence is expected from a theory for metallic and doped
semiconducting nanotubes \cite{[9],[13]}, and the positive sign
indicates hole-like major carrier \cite{[25]}. It is worth noting
that this observed TEP is markedly different from the previous
'bulk' measurement that exhibited somewhat smaller TEP and
deviations from a linear $T$ dependence \cite{[8]}. Again, we
believe that the numerous unknown tube-tube junctions in the 'mat'
sample may produce these differences, and that a mesoscopic
measurement of a single tube is necessary to properly probe the
intrinsic TEP of nanotubes.

In conclusion, we have presented, for the first time, mesoscopic
thermal transport measurements of carbon nanotubes. The observed
thermal conductivity of a MWNT is more than 3000 W/m K at room
temperature and the phonon mean free path is $\sim$ 500nm. The
temperature dependence of the thermal conductivity shows a peak at
320~K due to the onset of Umklapp phonon scattering. The
thermoelectric power shows an expected linear $T$ dependence,
which was absent previous bulk measurements. The experimental
techniques reported here should be readily applicable to other
nanoscale materials to study their thermal properties.

The authors wish to thank J. Hone, D. Li, S. Jhi, Y.-K. Kwon, and
D. Tomanek for helpful discussions. We also thank A. Rinzler and
R.E. Smalley for supplying the nanotube materials. LS and AM would
like to acknowledge the support of the DOE (Basic Energy Sciences,
Engineering Division).  PK and PLM were supported by DOE (Basic
Energy Sciences, Materials Sciences Division, the sp$^2$ Materials
Initiative).

\end{multicols}


\begin{references}
\bibitem{[17]}Below 8 K, both $R_s$ and $R_h$ become saturated due to the
impurity scattering, and cannot be used for a temperature sensor.
To insure that the measurement remains in the linear response
regime, $P$ was limited to make $T_h-T_0<1$ K during the
measurement.
\bibitem{[20]} The major uncertainty of $\kappa$ arises from the uncertainty in
diameter measurement. A high resolution SEM is used to determine
the diameter of MWNT to be 14 nm while the resolution limit is 2
nm.
\bibitem{[22]}The electrical contribution of thermal conductivity, $\kappa_{el}$,
can be estimated using Wiedemann-Franz law. From the
experimentally obtained electrical resistance of the MWNT,
$\kappa_{el} /\kappa \sim 10^{-3}$ at room temperature, and this
ratio becomes even smaller at the lower temperatures.
\bibitem{[24]} The thermal effects of one dimensional phonon quantization
in a nanotube should be measurable as $T<T_{1D}=h v/k_{B}R$, where
$h$ is the Plank constant, $k_B$ is the Boltzman constant, and $R$
is diameter of the nanotube. For a MWNT, $T_{1D}$ is estimated to
be $\sim$~1~K. Above this temperature, phonon transport in a MWNT
essentially behaves like in two dimensional graphene sheet.
\end{references}
\end{document}